# Reversible electric-field-driven magnetization in a columnar nanocomposite film

Mohsin Rafique[1,2,3,4], Andreas Herklotz[3], Kathrin Dörr[3], Sadia Manzoor[1,2,*]

[1]Magnetism Laboratory, Department of Physics, COMSATS University, 45550, Islamabad, Pakistan

[2]Center for Micro and Nano Devices (CMND), Department of Physics, COMSATS University, 45550, Islamabad, Pakistan

[3]Institute of Physics, MLU Halle-Wittenberg, 06099 Halle, Germany

[4]IFW Dresden, Postfach 270116, 01171 Dresden, Germany

*Corresponding author: sadia_manzoor@comsats.edu.pk

Abstract

Self-assembled heteroepitaxial $BiFeO_3 - CoFe_2O_4$ (BFO-CFO) nanocomposite films were grown on (001) oriented piezoelectric $Pb(Mg_{1/3}Nb_{2/3})_{0.72}Ti_{0.28}O_3(001)$ (PMN-PT) substrates by pulsed laser deposition in which BFO and CFO targets were alternately ablated. Completely reversible electric field induced biaxial strain in this substrate has been exploited to produce reversible electric field control of magnetization in the BFO-CFO nanocomposite film. The surface of the CFO-BFO films shows rectangular CFO pillars protruding out of a flat BFO matrix. The pillars are aligned with their edges along the pseudocubic {110} directions and are homogeneously distributed in the matrix. X-ray diffraction revealed an out-of-plane compression of the CFO unit cell by ~ 0.4%. Magnetic hysteresis loops show a moderate perpendicular anisotropy of the magnetostrictive CFO pillars, which is related to their vertical compression. The application of an electric field to the electromechanical PMN-PT substrate produced significant and reversible changes in the magnetization due to an additional strain-induced magnetic anisotropy. This demonstrates completely *reversible*, room-temperature electric-field-assisted control of magnetization in self-assembled vertical nanocomposites of CFO and BFO.

# 1. Introduction

Controlling magnetization directly with an electric field (converse magnetoelectric effect) is important for many different applications because it enables energy-efficient writing of magnetic bits as well as efficient control over magnetic permeability. It has applications in ultra-high density magnetic storage as well as in compact, miniaturized microwave devices [1, 2]. Composite multiferroics are known to exhibit strong, strain-mediated coupling between the electric and magnetic orders [1, 3-5]. There has been growing interest in coupling different functionalities of the magnetostrictive and piezoelectric components of these composites by interfacing them at the micro- or nanoscale [3, 6, 7]. This can be done by multilayering [8, 9], by growing vertical heteroepitaxial nanocomposites [4, 10-12], or by embedding particles of one phase in a three dimensional matrix of the other [13].

Perpendicular recording technology has been introduced in hard disk drives [14, 15] in order to overcome limitations posed by superparamagnetic effects [16] in traditional longitudinal recording media. Vertical nanostructures (pillar-matrix geometry) are interesting in this context because the magnetic nanopillars embedded in a nonmagnetic matrix have been shown to exhibit perpendicular magnetic anisotropy [4, 10, 11, 17]. Moreover, these vertical nanocomposites offer a large interfacial area between the magnetostrictive and piezoelectric phases as required for a significant magnetoelectric response [5]. This large interfacial area also provides an additional tool to control the elastic strain in the magnetic component, thus allowing tuning of the stress-induced magnetic anisotropy [5, 10, 11, 15, 18, 19]. Hence, the total magnetic anisotropy in columnar nanocomposites can be more sensitively tuned as in continuous films. Therefore, a large magnetoelectric response can be achieved in strain-optimized nanocomposites by a moderate additional electric-field-driven stimulus, for example, a piezoelectric substrate strain [11, 15, 20-24].

Among different electromechanical substrates, $Pb(Mg_{1/3}Nb_{2/3})_{1-x}Ti_xO_3$ has drawn huge interest in recent years. PMN-PT with x ~ 0.3 or 0.38 substrates have in-plane tetragonal domains. Wang et al. [15] and Tang et al. [25] have reported a giant magnetoelectric response in BFO-CFO nanocomposites grown on $Pb(Mg_{1/3}Nb_{2/3})_{0.7}Ti_{0.3}O_3$ and $Pb(Mg_{1/3}Nb_{2/3})_{0.62}Ti_{0.38}O_3$ substrates respectively, which arises because of reorientation of tetragonal domains in the substrate. These are difficult to switch reversibly with an applied electric field because of the large hysteresis involved [26]. Their findings [15, 25] have potential applications in heat-assisted magnetic recording devices where reversibility can be attained through a 'thermal reset', for example by annealing the electrically strained heterostructures at 200°C in air for 0.5 h [15]. *In contrast*, the present work is aimed at *reversible* electrical control of magnetization. This is made possible by using PMN-PT(001) with x = 0.28 [22, 24] as an electromechanical substrate. This particular composition of the substrate offers certain distinct advantages. Firstly, this composition is close to the morphotropic phase boundary on the rhombohedral side and offers a nearly cubic lattice [26-28] even in the ferroelectric state with a lattice parameter $a \sim 4.02$ Å and a rhombohedral angle of 89.9° at 300 K. Secondly, it does not exhibit any structural phase transitions [26] at and below room temperature, so that an electric field induced strain is almost independent of temperature. Lastly, weak hysteresis and time-dependent creep reported for extremely high strain levels of up to 0.6% [26] make it a versatile electromechanical substrate for electric field induced dynamic strain studies in films at and below 300 K. Our objective in using a PMNPT crystal with x = 0.28 was to apply a reversible biaxial strain [22, 24, 27] to the nanocomposite film, thus producing *reversible* control of the perpendicular magnetic anisotropy and hence the magnetization in the nanocomposite film.

## 2. Experimental Details

Two-phase self-assembled nanocomposite films can be grown by pulsed laser deposition in two different ways: i) ablating material from a single composite target [4, 29], which contains all the desired constituents of the film at or near their final average concentrations, and ii) alternatingly ablating materials from two different targets in the appropriate ratio [10-12, 18]. The latter approach offers a more flexible and precise control over film composition and quality [10-12, 18]. If the composite phases have weak mutual solubility, they will diffuse to separate out at a suitable growth temperature, forming nanopillars of one phase embedded in the matrix of the other [12, 30]. For the present study, nanocomposite $BiFeO_3$-$CoFe_2O_4$ (BFO-CFO) films were grown on $Pb(Mg_{1/3}Nb_{2/3})_{0.72}Ti_{0.28}O_3$ (PMN-PT) (001) substrates (*Atom Optics*) by alternately ablating two separate commercially obtained targets of $CoFe_2O_4$ and $BiFeO_3$ (Vinkarola Instruments, purity 99.999%). Growth temperature is a critical parameter for the separation of the two phases which ultimately determines the strain state of the constituent phases in the nanocomposite film [10, 18, 30]. In the present work, the nanocomposite film was grown at 650°C which is high enough for phase separation of BFO and CFO [30]. As shown in our previous work [11], BFO-CFO nanocomposite films grown at this temperature exhibit a moderate compressive strain of the CFO, which lends itself favorably to additional strain manipulation, for example by applied electric fields. Prior to the growth of the nanocomposite film, growth rates (0.02 nm/s for CFO and 0.1 nm/s for BFO) were calibrated for each constituent using X-ray reflectivity measurements on single-component films. The number of pulses for each component was chosen in such a way that the desired volume fraction of 40% CFO (columnar phase) and 60% of BFO (matrix phase) could be achieved. Before growing the nanocomposite film, a 20 nm thick $SrRuO_3$ (a = b = c = 0.393 nm) layer was grown on the PMN-PT substrate as a buffer layer. The chamber was evacuated to a base pressure of ~ $10^{-3}$ Pa before the introduction of oxygen into the chamber. Oxygen partial pressure of 10 Pa was used during deposition. The laser repetition rate was 8 Hz with a laser energy density of 1.2 J cm$^{-2}$. Samples were annealed post-growth at 650°C for 20 minutes in an oxygen pressure of 80 Pa in order to compensate oxygen vacancies arising during growth. The thickness of the composite films estimated from the total number of pulses is ∼ 200 nm. Magnetic and magnetoelectric properties were studied at room temperature using a superconducting quantum interference device (SQUID) magnetometer (*Quantum Design MPMS-5S*). A thin layer of conductive silver paint was used for the top and bottom electrodes. A schematic of the nanocolumnar composite film together with the top and bottom electrodes used for applying the voltage $V_{app}$ is illustrated in Fig. 1 (a).

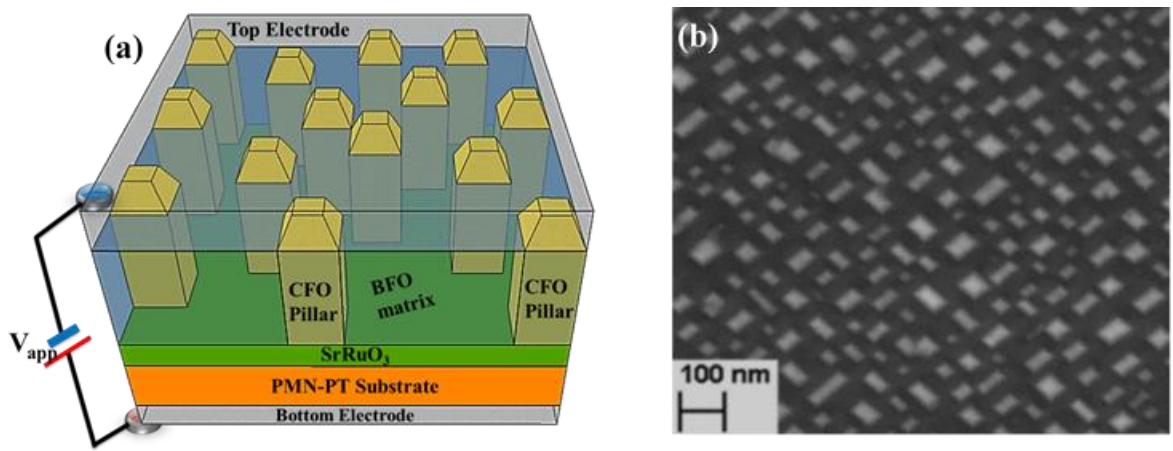

*Figure 1 (a) Schematic of the nanocolumnar composite film showing CFO nanopillars embedded in a BFO matrix grown on a PMN-PT substrate. Reversible strain can be produced in this structure by application of a voltage "$V_{app}$" between the top and bottom silver electrodes. (b) Scanning electron microscopy image showing the surface morphology of the nanocomposite BFO-CFO film with CFO islands protruding out of a flat BFO matrix.*

## 3. Results and Discussion

Fig. 1 (b) shows the surface morphology of the nanocomposite film as imaged by a thermally-assisted field emission scanning electron microscope (*FESEM, LEO 1530*). The surface shows well defined rectangular crystallites of CFO protruding out of a flat BFO matrix. Cobalt ferrite grows in the form of columnar grains in the BFO matrix, as reported earlier for the growth of this type of nanocomposite on a (001)-oriented perovskite-type substrate [12, 15, 17, 31]. Statistical analysis (not shown here, please see reference [18]) by considering 550 CFO islands, yields that rectangular islands of CFO have mean lengths and widths in the ranges 30 - 40 nm and 20 - 25 nm, respectively. This kind of morphology is a signature of 1-3 type nanocomposites with the CFO phase growing as nanopillars in a three-dimensional matrix of the perovskite phase as reported earlier [15] . Based on the cube-on-cube epitaxy of SRO and the BFO-CFO composite film on the substrate, the CFO crystallites are oriented with their edges along the pseudocubic {110} directions of the substrate [32]. Slutsker et al.[33-35] have explained the growth of this type of nanocomposites by assuming the different constituent phases present in the composite film each to be an elastic domain, with elastic interaction between themselves and the substrate surface. Under the constraint of a strain condition enforced by the substrate, elastic interaction of each phase with the substrate controls the morphology of the nanocomposites.

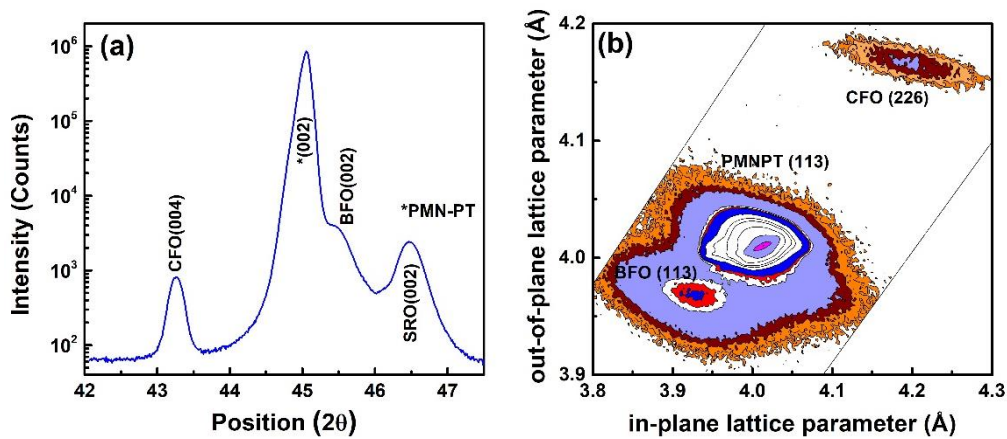

*Figure 2(a) High resolution Brag Brentano θ-2θ scan showing the 00l scan around the 002 peak of the PMN-PT substrate. (b) In-plane and out-of-plane lattice parameters calculated from the reciprocal space map around the (113) substrate reflections of the BFO-CFO nanocomposite film grown heteroepitaxially on a PMN-PT (001) substrate.*

The structural properties of the nanocomposite films were investigated by x-ray diffraction using a four-circle diffractometer (*Philips X'pert PW 3373*) with a Cu source ($\lambda$ = 1.54 Å). The high-resolution scan around the 002 peak of the substrate (Figure 2 a) resolved the CFO and SRO peaks. The BFO peak could not be completely resolved because of its proximity with the strong substrate peak as reported in literature [15], and appears as a shoulder on the right side of the substrate peak. The in-plane and out-of-plane lattice parameters of CFO and BFO have been determined from the reciprocal space map (RSM) around the (113) reflection of the PMN-PT substrate, as depicted in Fig 2 (b). CFO has a cubic structure with the lattice parameter a = b = c ~ 8.38 Å in bulk. RSM analysis shows that for CFO, the in-plane lattice parameter has a value of a ~ 8.38 Å, revealing relaxed growth at the interface with the SRO-buffered substrate because of the large lattice mismatch of ~ 6.6% with the latter. The out-of-plane lattice parameter of CFO has a value of c ~ 8.35 Å which is smaller than the bulk value and in agreement with calculations from Figure 2 (a). Thus, CFO is experiencing a compressive strain of ~ 0.4% in the vertical direction. BFO has a rhombohedrally distorted perovskite structure with $a_r$ ~ 3.96 Å and $α_r$ ~ 0.6° in bulk. Analysis of figure 2 (b) shows that the in-plane lattice parameter of BFO is a ~ 3.93 Å, smaller than the bulk value. The smaller unit cell of the SRO induces an in-plane compression of ~ 0.7 % in the BFO unit cell. The out-of-plane lattice parameter of BFO is c ~ 3.98 Å indicating a tensile strain of ~ 0.5 % in the vertical direction of the BFO unit cell. BFO´s elastic response (Poisson number) is known to be near 0.3 [24]. This means that vertical and in-plane strains have similar absolute values. The difference of 0.2% between the in-plane and out-of-plane strains could be due to measurement errors. These results can be explained as follows. The in-plane lattice parameters of the PMN-PT substrate at room temperature are a = b = 4.022 Å [26]. The SRO buffer layer grows epitaxially oriented, but strain-relaxed on the substrate with an out-of-plane lattice parameter of ~ 3.93 Å calculated from the high resolution θ-2θ scan around the (002) peak of substrate. SRO and BFO peaks could not be resolved for the in-plane lattice parameter in the reciprocal space map (Fig. 2b). Bulk BFO has a lattice parameter ($a_r$ ~ 3.96 Å) smaller than CFO (a/2 = 4.19 Å). The epitaxially grown vertical CFO/BFO interface induces a compression in the CFO unit cell. This is likely to induce a small tensile strain (~ 0.5%) in the BFO unit cell in the vertical direction. The small compression in the CFO unit cell (~ 0.4 %) shows that most of the vertical compression in the CFO phase has been released by formation of dislocations at the interfaces between BFO and CFO at the growth temperature [30]. These are expected to occur because of the large lattice mismatch of ~ 5.8% between BFO and CFO. The vertical strain induced in both of the constituent phases (compressive in CFO and tensile in BFO) because of the epitaxially grown interfaces, has been reported for similar nanocomposites grown on different substrates having (001) orientation, such as $BiFeO_3$-$CoFe_2O_4$ on $SrTiO_3$ [19], $BaTiO_3$ [11] and

Pb(Mg$_{1/3}$Nb$_{2/3}$)$_{0.7}$Ti$_{0.3}$O$_3$ [15]. The residual strain induced by epitaxy at vertical interfaces has drawn considerable attention [36] in these types of nanocomposites, since it provides a tool for controlling the properties of the constituent phases, such as magnetic anisotropy [4, 10-12, 18] and critical transition temperatures [37]. Since CFO is a well-established magnetostrictive material with a large negative magnetostriction coefficient $\lambda_{100} \sim -5.9 \times 10^{-6}$ [38], the small compression of ~ 0.4% in the CFO unit cell is expected to affect the magnetization hysteresis along both the pseudocubic 100$_{pc}$ in-plane direction of the film as well as along the film normal (out-of-plane direction). Room temperature M-H hysteresis loops measured in the in-plane and out-of-plane directions (Fig. 3 (a)) show a substantial remanence (M$_r$) for both orientations. The magnetization value has been normalized with respect to the CFO volume in the composite film estimated from the total number of pulses used to ablate the CFO target. The remanence ratio M$_r$ / M$_s$ is 60% for the out-of-plane magnetization and 35% for the in-plane measurement (saturation magnetization in both cases is taken to be 400 emu.cm$^{-3}$ [12]). This shows that the nanocomposite film exhibits moderate perpendicular magnetic anisotropy.

To identify the source of this anisotropy, we calculated the anisotropy field after taking into account the different sources which contribute to the total energy of the film (for details, see references [12, 18]). These sources include Zeeman energy, shape anisotropy, bulk magnetocrystalline anisotropy energy and magnetoelastic energy. The calculated values of the anisotropy fields arising because of vertical compression (~ 0.4%) and shape (considering CFO nanopillars with an aspect ratio of ~5 are ~ 2.1 T and ~ 0.2 T, respectively. So it can be concluded that magnetic anisotropy arises primarily because of the compression along the c-axis of the CFO nanopillars (as seen in the (002) scan) in conjunction with the large negative magnetostriction of CFO. Figure 3 (a) shows that our sample exhibits a moderate perpendicular magnetic anisotropy in contrast to the observed strong anisotropy reported in our earlier work [10] and in the work of some other authors [12, 15]. The weakness of the observed perpendicular anisotropy can be attributed to the small vertical compression of the CFO nanocolumns [10, 12]. This is also the reason behind the small coercive field ~ 0.5 T (out-of-plane), in contrast to values beyond 1T observed in strongly compressed CFO columns [10, 12, 15].

Moderate perpendicular anisotropy resulting from the weak 'as-grown' strain in the CFO nanopillars is advantageous in this study over strong magnetic anisotropy because it allows us to modulate the magnetic properties of the film using a small strain, driven by an electric field applied across the thickness of the sample. We note here that the magnetic response of the nanocomposite film is entirely due to the CFO nanopillars. Bulk BFO is antiferromagnetic at room temperature ($T_N$ = 643 K) and though ferromagnetism has been reported for very thin BFO films [39], it is expected to be negligible for our sample with film thickness ~ 200 nm. An electric field of 7.5 kV/cm was induced along the substrate normal by applying a voltage of 300 V between the top and bottom silver paint electrodes sandwiching the nanocomposite film as shown in the schematic diagram of Fig. 1 (a).

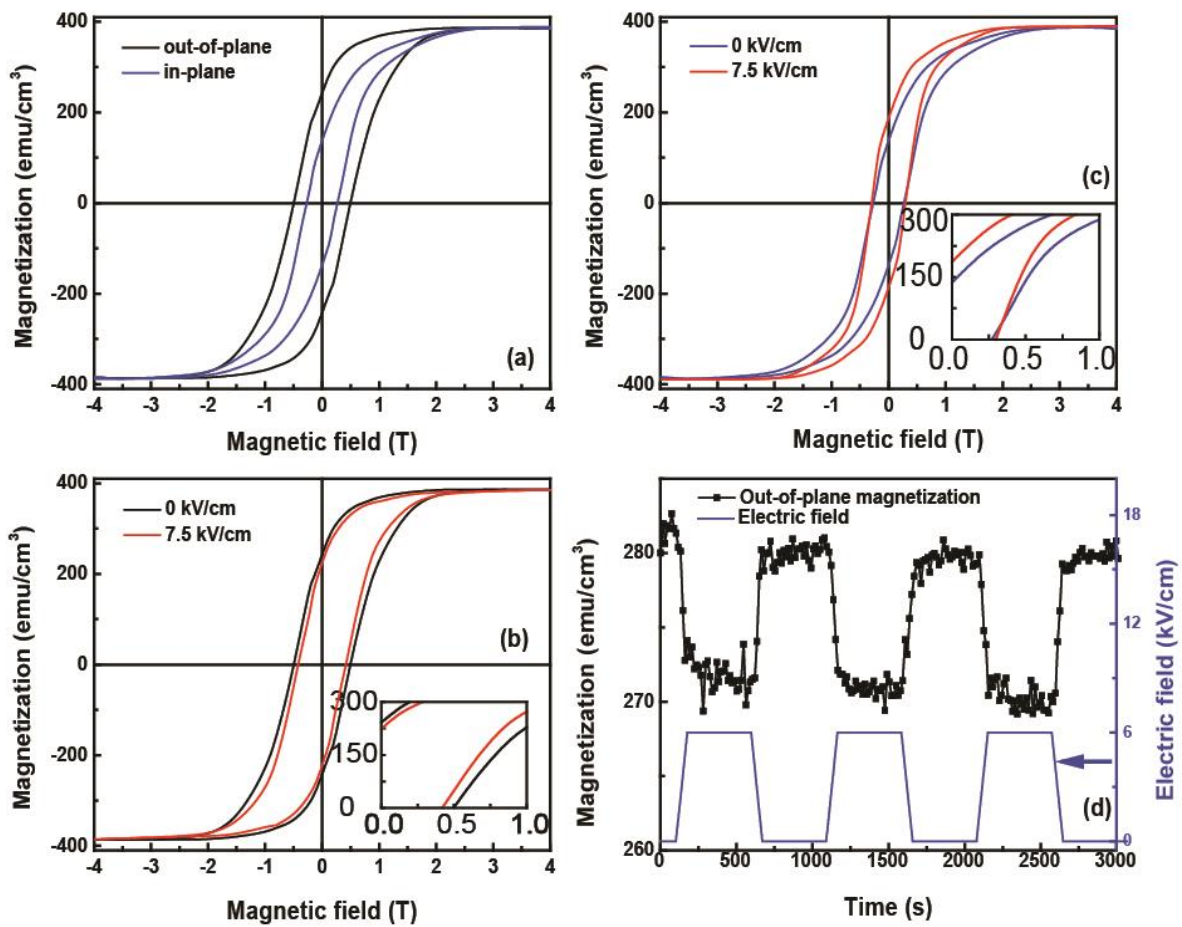

*Figure 3 Magnetization hysteresis loops for the BFO-CFO nanocomposites grown on a (001) oriented PMN-PT substrate, measured at room temperature (a) out-of-plane (black) and in-plane (blue). In-plane (b), and out-of-plane (c) M-H loops measured with electric field E = 0 (black) and E = 7.5 kV/cm (red). Insets show remnant magnetization. (d) Out-of-plane magnetic response to a periodic electric field (continuous orange line) measured with a magnetic bias field of 0.1 T.*

Biegalski *et. al* [24] et al. have shown that application of an electric field up to 13.3 kV/cm applied across the thickness of a Pb(Mg$_{1/3}$Nb$_{2/3}$)$_{1-x}$Ti$_x$O$_3$ (001) crystal (with $x$ = 0.28) results in a linear decrease in the in-plane lattice parameters 'a' and 'b' producing a maximum compression of 0.13 ± 0.01 % and a corresponding linear increase in the out-of-plane lattice parameter 'c' of 0.20 ± 0.001%. Moreover, these observed changes in the unit cell of the PMN-PT (x = 0.28) crystal are completely reversible with applied field [24]. Thus an electric field of 7.5 kV/cm applied to the substrate allows one to reversibly reduce the biaxial in-plane lattice parameter of the substrate by 0.08%, see Ref. [24]). This produces an equal in-plane compression in any film grown epitaxially on top of it [22, 24, 40]. Due to this electric field induced biaxial in-plane compression, the BFO matrix experiences an elongation in the out-of-plane orientation. As a result, the as-grown moderate compression in the CFO unit cell decreases and hence the strain controlled magnetic properties of CFO nanopillars are modified [11, 15, 18]. The electric field induced reduction in the compressive strain of CFO is facilitated by the BFO matrix which acts as a bridge [11, 12, 15] for the strain transfer between the substrate and the CFO pillars.

Considering the negative magnetostriction of CFO, one expects that the in-plane compression and out-of-plane relaxation of the nanocomposite film induced by the electric field, would produce an enhancement of the in-plane magnetization and a reduction in the out-of-plane magnetization, respectively, in the hysteretic range of the magnetization loop. This can be seen clearly in Figs. 3 (b) and (c), which show the in-plane and out-of-plane M-H loops with electric field E = 0 and E = 7.5 kV/cm. Under application of the electric field, the out-of-plane remanence $M_r^{oop}$ decreases by ~ 10% (see inset), while the strain-induced magnetization change close to magnetic saturation almost vanishes for both measuring directions. The remanence enhancement in the in-plane direction is larger (~ 40%) (see inset). The calculated magnetoelectric coefficients in the in-plane and out-of-plane orientations using the changes in the remanence magnetization have values of ~ $7 \times 10^{-8}$ s m$^{-1}$ and ~ $2.6 \times 10^{-8}$ s m$^{-1}$, respectively. The coercivity of the nanocomposite film is also influenced by the electric field induced elastic strain when measured along the column axis. The effect of the electric field is stronger for the out-of-plane coercive field than for the in-plane value which reveals comparatively little reaction to the reversible strain change (Fig. 3b and c).

Fig. 3d) shows an example of reversible switching at room temperature of the out-of-plane magnetization with an applied voltage of 300 V corresponding to an electric field E = 6 kV/cm, in the presence of a bias magnetic field of 0.1 T. From Fig. 3d), it can be seen that the remnant magnetization increases by about 4% under application of an electric field of 6 kV/cm. This change is reversible with cycling of the electric field. The reversibility observed in Fig. 3d) also exists even in the absence of a bias magnetic field (not shown here). These results show that the in-plane magnetization of the BFO-CFO composite film follows the electric field applied to the PMN-PT substrate in a reversible manner at room temperature.

We could not find a report on electric field driven reversible strain control of magnetization in BFO-CFO nanocomposites exhibiting perpendicular magnetic anisotropy grown on PMN-PT (x = 0.28). In our previous work [11], we have demonstrated a giant magnetoelectric effect in a bismuth ferrite (BFO)-cobalt ferrite (CFO) nanocolumnar composite film on a ferroelectric multi-domain BaTiO3(001) substrate. Wang et al. [15] and Tang et al. [25] have also reported a giant magnetoelectric response in BFO-CFO nanocomposites grown on Pb(Mg1/3Nb2/3)0.7Ti0.3O3 and Pb(Mg1/3Nb2/3)0.62Ti0.38O3 substrates respectively. The above-mentioned substrates have tetragonal unit cells, with electric dipoles oriented along the c axis of the unit cell making it difficult to achieve reversibility with electric field. In all of these reports, the large magnetoelectric response appears because of re-orientation of in-plane tetragonal domains in the substrate upon application of an electric field along the substrate normal. This response was found to be reversible only with thermal treatment at a temperature close to the Curie temperature of the substrates. We have achieved reversibility of the magnetoelectric response by using PMN-PT crystals with x = 0.28. These have monoclinic structure with the ferroelectric polarization P lying along the body diagonal of the pseudocubic unit cell. This induces precisely controlled, reversible and uniform in-plane strain in the epitaxially grown film under an applied electric field [24]. This feature of the PMN-PT crystal has been exploited in this work to achieve completely reversible electric field assisted control of magnetization at room temperature. The present study is significantly different from those mentioned above [11] in that it eliminates the need of thermally 're-setting' the ferroelectric domains. The reversibility demonstrated in our experiment makes these systems very attractive for magnetoelectric device applications. In contrast to thin films grown epitaxially on a substrate [20], nanocomposites with pillars of CFO embedded in the ferroelectric matrix are in general attractive as none of the phases is completely clamped by the substrate and they possess perpendicular magnetic anisotropy. Epitaxially grown vertical interfaces can introduce very large strains even in much thicker films (~ μm) [10, 12], and act as a bridge to transfer strain from one phase to the other giving a large ME response.

4. **Summary and Conclusions**

In summary, magnetostrictive CFO pillars in self-assembled nanocomposites grown on (001) oriented PMN-PT substrates experience an out-of-plane compression of about 0.4% resulting in perpendicular anisotropy. PMN-PT (001) substrates with x = 0.28 are useful in producing completely reversible electric control of magnetization in the epitaxially grown nanocomposite films. The electric field induced changes in magnetization are the consequence of elastic coupling between the piezoelectric and piezomagnetic phases. Using the changes in the remanence, the calculated magnetoelectric coefficients in the in-plane and out-of-plane orientations have values of ~ $7 \times 10^{-8}$ s m$^{-1}$ and ~ $2.6 \times 10^{-8}$ s m$^{-1}$, respectively. Although these represent a relatively weak magnetoelectric coupling, the demonstrated reversibility of magnetization changes under applied electric field is remarkable and paves the way for further investigation to enhance the magnetoelctric response. This may be achieved by exploring different matrix/nanopillar materials as well as improving the nature of the interfaces. Reversibility is important in the context of electric field control of data writing and can be useful in overcoming the power dissipation issue in the present current driven write technology and microwave devices. From the point of view of applications, a more uniform size distribution and controlled positions of the CFO columnar grains would be desirable. This can be achieved by using pattering techniques [31, 41] in combination with self-assembly.

**Acknowledgement**

The Higher Education Commission (HEC), Govt. of Pakistan and Deutsche Forschungsgemeinschaft (DFG) are highly acknowledged for providing financial support. HEC supported under lab grant (No. 1-28/Lab/COMSATS/Acad-R/HEC/2010/933) and funded the author, Mohsin Rafique, under the Indigenous Programme (PIN: 074-29 71-Ps4-477) and International Research Support Initiative Programme (PIN: IRSIP 20 PS 21). While working as a guest scientist, he received financial support from the SFB 762 (project A9) funded by DFG.